\documentclass[conference]{IEEEtran}
\IEEEoverridecommandlockouts
\usepackage{url}
\usepackage{cite}
\usepackage{booktabs}
\usepackage{amsmath,amssymb,amsfonts}
\usepackage{algorithmic}
\usepackage{graphicx}
\usepackage{textcomp}
\usepackage{xcolor}
\def\BibTeX{{\rm B\kern-.05em{\sc i\kern-.025em b}\kern-.08em
    T\kern-.1667em\lower.7ex\hbox{E}\kern-.125emX}}
    
\begin{document}

\title{PitchFlower: A Flow-based Neural Audio Codec \\ with Pitch Controllability}

\author{
\IEEEauthorblockN{Diego Torres, Axel Roebel, Nicolas Obin}
\vspace{5pt}
\IEEEauthorblockA{STMS Lab \\
IRCAM - CNRS - Sorbonne Université\\
Paris, France}
}

\maketitle

\begin{abstract}
We present PitchFlower, a flow-based neural audio codec with explicit pitch controllability. Our approach promotes pitch disentanglement through a simple perturbation: during training, F0 contours are flattened and randomly shifted at the input, while the true F0 is provided as conditioning to regenerate the original audio. A vector-quantization bottleneck prevents pitch recovery, and a flow-based decoder generates high quality audio. Experiments show that PitchFlower achieves accurate pitch control at the level of DSP baselines but at much higher audio quality, and performs on par with state-of-the-art neural approaches. Notably, despite using WORLD-transformed audio for training, our method filters out the vocoder's inherent artifacts, revealing a strong resilience of deep generative modeling to input degradation. This finding suggests that our framework provides a simple and extensible path that could be extended to other speech attributes.

\end{abstract}


\section{Introduction}

Speech attribute manipulation has long been an active area of research, with the goal of enabling modifications of interpretable features such as emotion, accent, speaker identity, or pitch. Pitch control, in particular, allows speech and music to be altered in meaningful ways. Its most familiar application is vocal tuning, where the singer’s pitch must match the intended note. In speech, changing prosody (of which pitch is a fundamental part) directly affects meaning and communicative intent.
Traditional approaches to pitch control are rooted in the source-filter model of voice production. For example, the vocoder WORLD \cite{world} decomposes speech into F0, spectral envelope, and aperiodicity ratio. Pitch can then be modified by adjusting the extracted F0 and resynthesizing the signal. However, such manipulations ignore the interactions between F0 and the other speech attributes, often resulting in unnatural-sounding audio with noticeable artifacts. 
More recent methods leverage deep learning \cite{frederik, sifigan, periodgrad, diff-pitcher, fastvgan, karatuner, promonet, neurodyne} to achieve pitch control. Some, such as SiFiGAN \cite{sifigan}, still embed assumptions from the source-filter model. Others apply them more indirectly, working with WORLD-derived representations as in PeriodGrad \cite{periodgrad}. Alternative strategies include conditioning on explicit, interpretable attributes (e.g., FastVGAN \cite{fastvgan}) or combining learned and explicit representations, as in Promonet \cite{promonet} and KaraTuner \cite{karatuner}. These examples highlight how the choice of representation is tightly coupled with the feasibility of pitch control.

In parallel, neural audio codecs (NACs) have emerged as a powerful representation of speech. By leveraging neural networks and large-scale data, NACs achieve higher fidelity at lower bitrates than traditional codecs \cite{soundstream, encodec, dac}. Crucially, they have also become a backbone for generative tasks such as text-to-speech, positioning NACs as a standard representation for modern speech and audio systems.  
This has naturally raised the question of disentanglement within NACs. Most prior work has focused on separating linguistic content from speaker identity, with SpeechTokenizer \cite{speechtokenizer} being a notable example: it leverages self-supervised models to impose a dedicated semantic level within the codec. Pitch disentanglement, however, has received far less attention. FACodec \cite{naturalspeech3} considers four factors simultaneously (content, speaker, prosody, and acoustic details) but its high-level prosody codes do not allow for precise F0 control. PeriodCodec \cite{periodcodec}, designed for singing voice, is to our knowledge the first NAC with explicit pitch control. However, it inherits training instabilities of other GAN-based codecs and introduces additional losses and hyperparameters.  \\

We introduce \textit{PitchFlower}\footnote{Not to be confused with \textit{PitchFlow} \cite{pitchflow}, which is a TTS and uses a radically different approach.}, a neural audio codec with precise pitch controllability. PitchFlower encourages pitch disentanglement through a simple perturbation-based strategy: we perturb pitch information at the input and task the model to reconstruct the original mel-spectrogram, conditioned on the ground-truth F0. 



\begin{figure}[htb]
    \begin{minipage}[b]{1.0\linewidth}
      \centering
      \centerline{\includegraphics[width=0.95\linewidth]{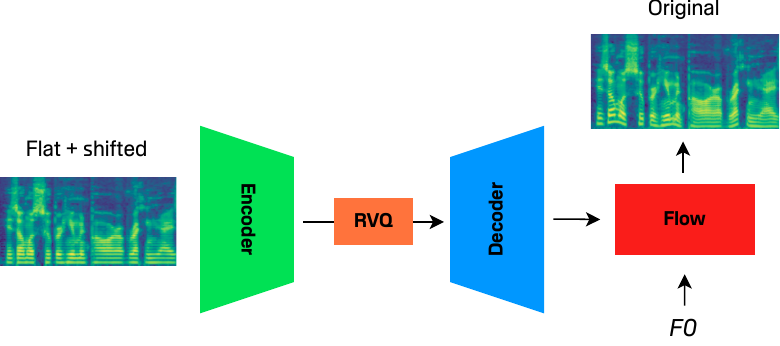}}
  \end{minipage}
  \caption{{\em Architecture and training methodology of PitchFlower. Input audio is perturbed through a flattening and random shifting of its F0 contour. An autoencoder derives hidden features that serve as conditioning for a flow-matching module, which uses them along the ground-truth F0 to reconstructe the original audio.}}
  \label{fig:architecture}
  \vspace{0.25cm}
\end{figure} 

The main contributions of this work are:  
\begin{itemize}
    \item We propose a perturbation+bottleneck methodology that enforces disentanglement without additional modules or losses. This allows PitchFlower to be trained with a simple generative loss. 
    \item We provide a systematic comparison of disentanglement strategies (bottleneck, adversarial, semantic distillation), analyzing their trade-offs in terms of controllability, audio quality, and information preservation.  
    \item We demonstrate that PitchFlower achieves a degree of controllability equivalent to DSP-based baselines at much higher audio quality, and competitive overall performance with state-of-the-art neural approaches.  \footnote{We encourage readers to listen to our demo page, at \url{https://pitchflowerdemo.account-229.workers.dev/}}
\end{itemize}

\section{PitchFlower}
\label{sec:pitchflower} 

PitchFlower adopts the standard architecture of recent flow-based audio codecs \cite{flowdec, flowmac}, consisting of an autoencoder, a vector-quantization bottleneck, and a flow decoder (Figure~\ref{fig:architecture}). The F0 contour is provided as an explicit conditioning signal to the flow decoder. The objective is to disentangle pitch such that latent representations are free of F0 information, allowing pitch to be directly controlled by modifying the conditioning contour.

To enforce disentanglement, we perturb the input during training by flattening its F0 contour. Specifically, each frame’s F0 is replaced with the utterance-level mean plus a random shift sampled from $\mathcal{U}(-\Delta,\Delta)$. WORLD is used to extract and modify the F0 values for this transformation. 

A key element of PitchFlower is the flow decoder. Since perturbation and quantization inevitably remove information beyond pitch, the flow decoder compensates by sampling plausible values from the learned distribution, ensuring realistic and high-quality audio. Formally, given a perturbed mel-spectrogram with autoencoder output $e$, a target mel-spectrogram $x_1$, a noise sample $x_0 \sim \mathcal{N}(0,1)$, and the flow decoder $v_\theta$. The conditional flow-matching loss is then \cite{flowmatching}:

\begin{equation}
    \mathcal{L}_{\text{CFM}}
    =
    \mathbb{E}
    \left\lVert
    v_\theta\,\big(x_t,\, e, f_0,\, t\big)
    - \Big(x_1 - (1 - \sigma_{\min}) x_0\Big)
    \right\rVert^2.
\end{equation}

Unlike prior work \cite{flowdec, flowmac}, where the flow acts as a post-net refining autoencoder outputs, here everything is trained end-to-end with the flow-matching loss as the only generative loss. 

The final loss is then
\begin{equation} \mathcal{L} = \mathcal{L}_{\text{CFM}} + \lambda_{\text{commit}}\,\mathcal{L}_{\text{commit}}.
\end{equation}

\vspace{0.5cm}

\section{Experimental Framework}

\subsection{Dataset and Implementation Details}
We train and evaluate our model on the LibriTTS dataset \cite{libritts}. The entire dataset is preprocessed offline, as opposed to an on-the-fly approach. The pitch perturbation range is set to $\Delta=5$ semitones.
The encoder and decoder are built from two ConvNeXt blocks \cite{convnext} (6 layers each), followed by self-attention \cite{attention}. The second block uses a stride of 2 for down/up-sampling. The RVQ module contains 8 codebooks with 512 entries each (dimension 256). The flow decoder consists of 4 blocks with 8 layers each. Hidden dimensions are 512 for the autoencoder and 256 for the flow module. The F0 contour is encoded with a 3-layer MLP (64 units per layer), with separate embeddings for unvoiced frames and missing F0.
For training, we adopt conditional flow matching \cite{flowmatching} with $\sigma_{\min}=10^{-4}$. Classifier-free guidance (CFG) \cite{cfg_flow} is applied by dropping either conditioning (F0 or decoder outputs) 10\% of the time.

The model is trained for 1M iterations on a single RTX 4070 GPU, using AdamW (lr=$10^{-4}$), batch size 32, and 1.5-second audio segments. The commitment loss weight is set to $0.25$. Vocos \cite{vocos} is used as vocoder to produce waveforms from mel-spectrograms.

\subsection{Evaluation Metrics}

We design our evaluation around the basic requirements for a pitch-controllable model: accurate modification of the F0 with high audio quality. In addition, the model should preserve the linguistic content of the utterance and minimize changes to the speaker’s voice quality. To capture these aspects, we rely on four objective metrics, evaluated on the dev-clean subset of LibriTTS with pitch shifts ranging from $-6$ to $+6$ semitones.  

\textit{(1) Word error rate.} We use an ASR model to transcribe the transposed audio and compare it to the original transcription. Specifically, we use the English-only medium-sized \footnote{https://huggingface.co/openai/whisper-medium.en} version of Whisper \cite{whisper}.  

\textit{(2) Speaker similarity.} We measure similarity between the original and transposed audio using speaker embeddings extracted with ECAPA-TDNN \cite{ecapatdnn}\footnote{https://huggingface.co/speechbrain/spkrec-ecapa-voxceleb}, compared via cosine similarity.  

\textit{(3) F0RMSE.} We compute F0 contours for both the original and transposed audio. The original contour is shifted by the target amount, and the root mean square error (Hz) is computed against the transposed contour. F0 estimation is performed with CREPE \cite{crepe}\footnote{https://github.com/maxrmorrison/torchcrepe}.  

\textit{(4) UTMOS.} We use UTMOS \cite{utmos}\footnote{https://github.com/sarulab-speech/UTMOS22} to automatically estimate the perceptual quality of the transposed audio.

\section{Comparison with related work}
\label{sec:comparison}

\subsection{Objective evaluations}
We compare PitchFlower against two established methods: WORLD \cite{world}, a DSP source–filter vocoder, and SiFiGAN \cite{sifigan}, a neural vocoder with pitch controllability. We note that although more recent approaches such as PeriodCodec \cite{periodcodec} or Neurodyne \cite{neurodyne} have been released, they are not publicly available. SiFiGAN remains a closely related method, having in common both the GAN-based approach and the source excitation. We also include a baseline where WORLD-transposed audio is resynthesized using Vocos \cite{vocos}, which acts as a post-processing module intended to remove artifacts and increase quality. In addition, we evaluate a variant of our model, PitchFlowerUV, where the pitch perturbation follows DiffPitcher \cite{diff-pitcher}: instead of the flat+shift operation, all F0 values are replaced with unvoiced frames before being resynthesized with WORLD. For the PitchFlower models, we use a CFG guidance scale of 3.0 for both conditionings, which we found to give better results.

\begin{figure}[htb]
    \begin{minipage}[b]{\linewidth}
      \centering
       \centerline{\includegraphics[width=1.0\linewidth]{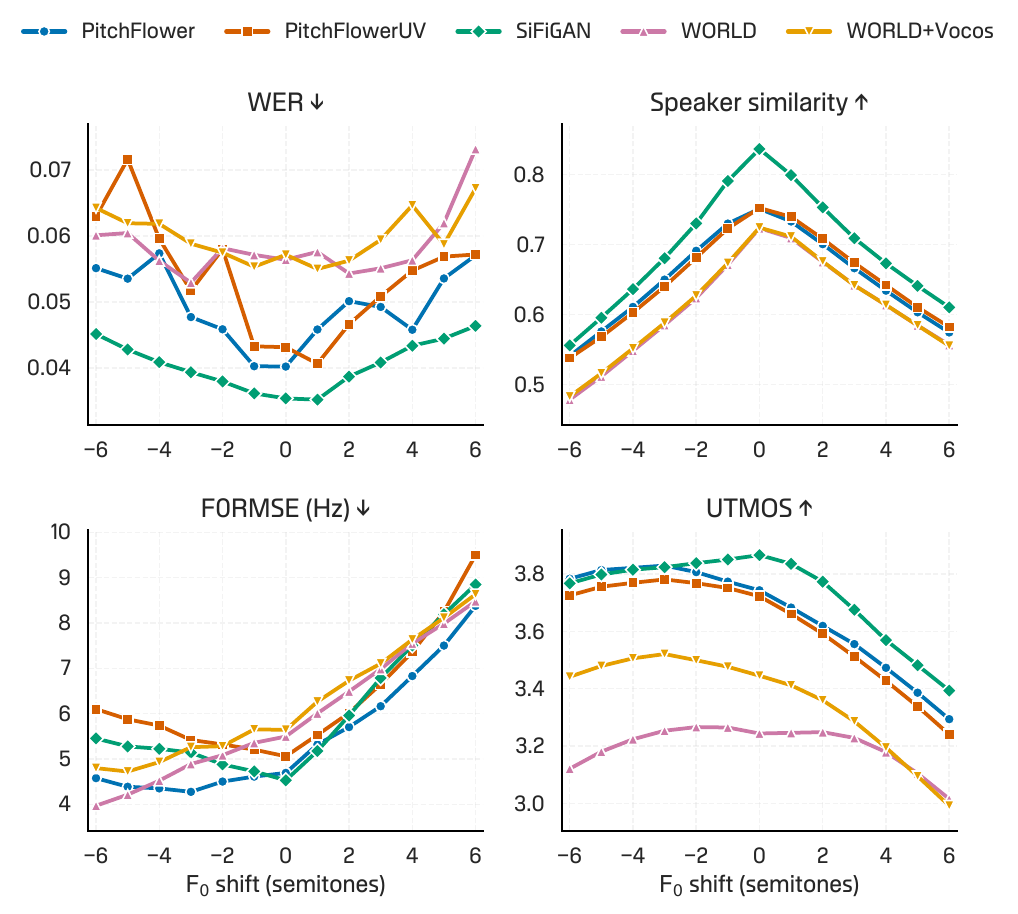}}
  \end{minipage}
  \vspace{-0.3cm}
  \caption{Objective evaluation results comparing PitchFlower with baselines. An alternative version of our model, PitchFlowerUV, is also considered.}
  \label{fig:comparison}
\end{figure}

Figure~\ref{fig:comparison} shows the objective evaluation results for the five methods. Regarding pitch controllability, PitchFlower achieves the strongest performance, as evidenced by its lower F0RMSE curve. For audio quality, a clear difference can be seen between WORLD and the deep learning-based methods, with the former having a considerably lower UTMOS score than its neural counterparts. Adding Vocos to WORLD has a positive effect on audio quality, but it is not enough to clear the gap with the other methods. 
As for intelligibility (WER) and speaker similarity, SiFiGAN has the best performance, followed by PitchFlower, PitchFlowerUV and finally WORLD. We note that the higher WER of PitchFlowerUV with respect to PitchFlower can be explained by the total unvoicing operation, which destroys the underlying harmonic structure and neutralizes phonemic distinctions between voiced-unvoiced pairs such as /s/–/z/ or /f/–/v/. In contrast, PitchFlower’s flat+shift perturbation merely strips the specific F0 trajectory while preserving these essential phonetic voicing traits, allowing the encoder to retain cleaner, more robust linguistic representations.

Notably, despite being trained on audio processed by WORLD, PitchFlower successfully filters out the vocoder's inherent artifacts. PitchFlower consistently outperforms WORLD across audio quality, speaker similarity, and intelligibility metrics. This superiority indicates that the flow decoder does not merely replicate the degraded target distribution but instead learns a clean generative prior, acting as an implicit denoiser. This capability goes far beyond the naive post-processing approach of pairing WORLD with Vocos, which yields only marginal improvements in audio quality while leaving speaker similarity and intelligibility entirely unimproved. Furthermore, experiments indicate a slight advantage for PitchFlower in pitch controllability. This performance gain is likely due to the reduction of artifacts, which facilitates more accurate pitch extraction for the CREPE algorithm \cite{crepe}.

\subsection{Subjective evaluations}

Based on the objective results, we conducted subjective evaluations focusing on the two top-performing methods: PitchFlower and SiFiGAN. We employed a Comparative Mean Opinion Score (CMOS) test to determine which model yielded superior results in general audio quality and speaker similarity.

For audio quality, participants were presented with pairs of randomly sampled audio processed by both methods and asked to identify which version exhibited higher quality. A similar approach was used for speaker similarity, with the addition of the original reference audio to provide participants with a ground-truth baseline for comparison. Following standard procedure \cite{cmos_citation}, we employed a CMOS scale ranging from -3 to +3, corresponding to "Much Worse" through "Much Better" ratings. This subjective evaluation was applied to audio samples whose pitch had been transposed by -6, 0, and +6 semitones. Evaluation was conducted with 25 participants, native English speakers, found through the Prolific platform. Each participant performed 20 evaluations, 10 of quality and 10 of similarity. We used the bootstrapping method to derive 95\% confidence intervals for the mean of the CMOS distribution.

\begin{figure}
    \centering
    \includegraphics[width=0.9\linewidth]{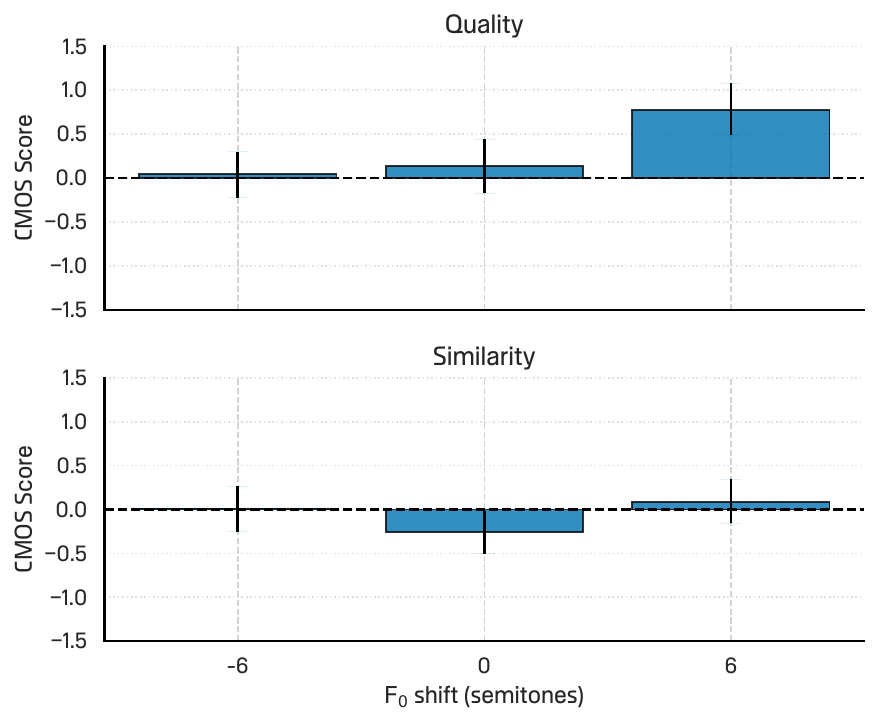}
    \caption{Subjective evaluation results. PitchFlower vs SiFiGAN. A positive value indicates preference for PitchFlower, a negative value preference for SiFiGAN. We show the mean CMOS score and its 95\% confidence interval.}
    \label{fig:cmos_results}
\end{figure}
\begin{figure*}[htb]
    \begin{minipage}[b]{1.0\linewidth}
      \centering
       \centerline{\includegraphics[width=1.0\linewidth]{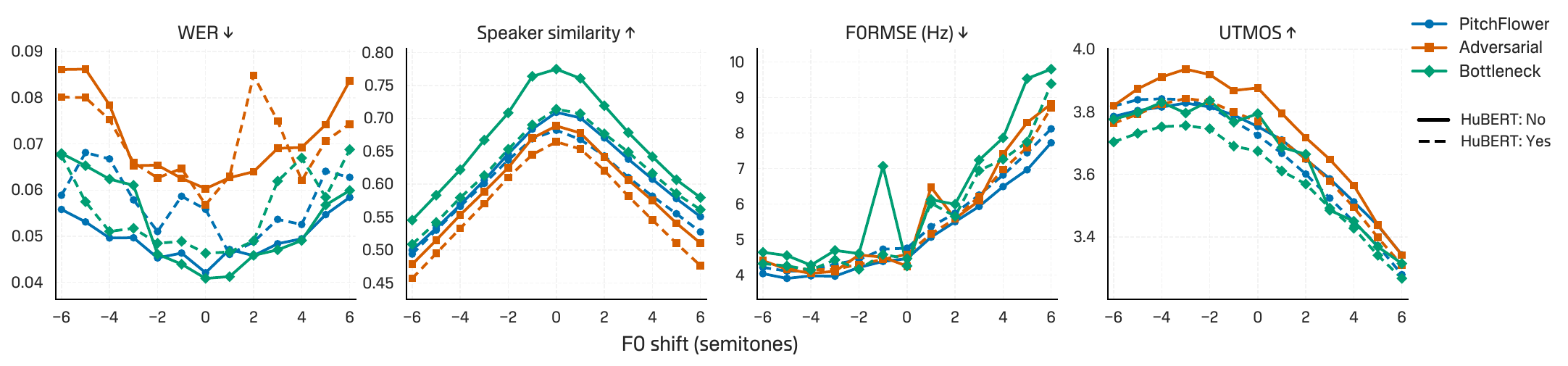}}
  \end{minipage}
  \caption{Objective comparison of different disentanglement strategies. Line colors indicate the main method used, and the dashed or continuous line indicates whether HuBERT distillation is used.}
  \label{fig:disentanglement}
\end{figure*}
As shown in Figure~\ref{fig:cmos_results}, both models have very close performances. PitchFlower achieves significantly higher audio quality for audios transposed +6 semitones, with a confidence interval for the CMOS mean of (0.48, 1.07). SiFiGAN achieves better similarity for reconstructed audios (0 semitones), with a confidence interval of (-0.50, -0.01). In the other configurations, despite finding a slight preference for PitchFlower in the CMOS means, no significant difference is found. We note that similarity results closely follow the objective evaluations, with SiFiGAN being superior in reconstruction but equivalent when performing big transpositions. We can attribute this behavior to PitchFlower's training procedure, which always has some pitch transposition involved.

\section{Ablation Studies}
\label{sec:ablation}

\subsection{Disentanglement Strategies}
\label{sub:disentanglement}
We investigate several strategies for disentanglement within the same framework. Unlike PitchFlower, these variants do not apply F0 masking at the input, and instead rely on alternative mechanisms to separate pitch information.  

\textit{(0) Bottleneck-based disentanglement.} Even without explicit masking, the RVQ module acts as an information bottleneck, encouraging the model to exclude F0 from the codes and rely on the conditioning signal instead. This effect alone provides a non-trivial degree of controllability, following the principle first exploited in AutoVC \cite{autovc}.  

\textit{(1) Adversarial disentanglement.} Inspired by PeriodCodec \cite{periodcodec} and NaturalSpeech3 \cite{naturalspeech3}, we add a pitch predictor with a gradient reversal layer. The predictor takes the quantized bottleneck as input and classifies log-F0 into one of six bins: five bins spanning 32–1024 Hz (log-scale) plus one for unvoiced frames. The gradient reversal layer encourages the encoder to remove pitch information from the latent variables. 

\textit{(2) Semantic distillation.} Following the line of SpeechTokenizer \cite{speechtokenizer}, we apply a distillation loss from HuBERT \cite{hubert} onto the autoencoder representations. Unlike prior work that constrains only the first quantization level, we distill the full RVQ output. The loss is computed as cosine similarity between HuBERT features and the corresponding RVQ states.  

Combining these components yields six variants for comparison: (i) bottleneck-only, (ii) bottleneck + HuBERT distillation, (iii) adversarial, (iv) adversarial + HuBERT, (v) PitchFlower, and (vi) PitchFlower + HuBERT.

Results are summarized in Figure~\ref{fig:disentanglement}. They can be analyzed along two dimensions: the effect of semantic distillation with HuBERT, and the comparison between disentanglement strategies.  

\textit{Effect of HuBERT.} Adding HuBERT generally degrades intelligibility and speaker similarity. For both PitchFlower and the bottleneck baseline, WER increased when the distillation loss was applied, while the adversarial variant showed mixed results. Speaker similarity and UTMOS scores consistently dropped across all methods. The only clear benefit of HuBERT was improved F0 controllability, as seen in lower F0RMSE for the bottleneck and adversarial methods. This suggests that HuBERT encourages more pitch--independent representations, but at the cost of other aspects of signal quality. The exception is PitchFlower, where HuBERT was detrimental across all metrics.   

\textit{Comparison of methods.} Among the three base strategies, the adversarial method achieved the worst WER and speaker similarity. This effect likely arises because the encoder, when forced to hide pitch information from the predictor, may also suppress other speaker-related cues, leading to degraded intelligibility and voice preservation. Interestingly, it yielded the highest UTMOS scores. PitchFlower and the bottleneck baseline were more balanced, showing comparable intelligibility and audio quality. However, PitchFlower exhibited lower speaker similarity, which we attribute to distortions introduced by WORLD during pitch perturbation.  In terms of disentanglement, PitchFlower provided the most accurate F0 control, followed by the adversarial method, with the bottleneck baseline last. Adding HuBERT improved the baseline to match the adversarial method, and combining adversarial training with HuBERT further stabilized performance by reducing large peaks in the F0RMSE curve.  

\subsection{Effect of Bottleneck}

\begin{figure}[htb]
    \begin{minipage}[b]{\linewidth}
      \centering
       \centerline{\includegraphics[width=0.8\linewidth]{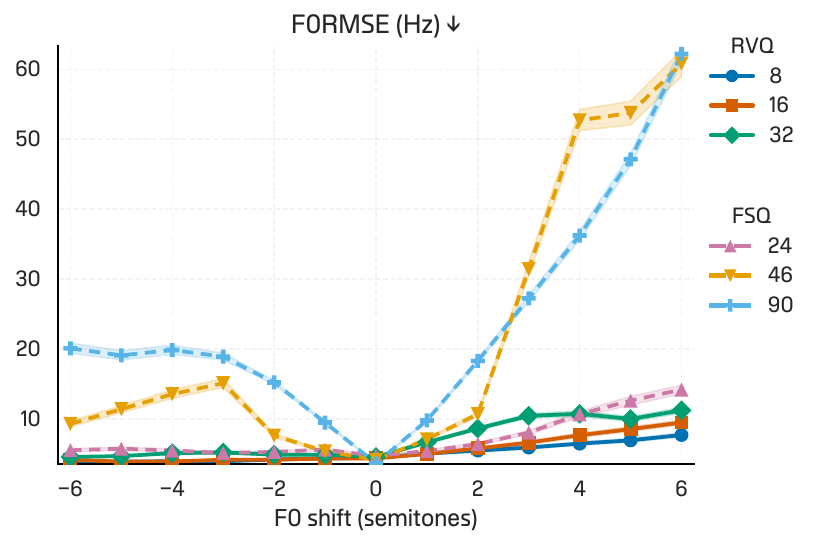}}
  \end{minipage}
  \caption{Pitch control for different sizes and types of bottleneck. A quick degradation is observed for the FSQ model as the size of the bottleneck increases. The RVQ variant is more robust to this variable. We compare methods with similar information capacity.}
  \label{fig:rvq_fsq}
\end{figure}

We analyze how the type and size of the bottleneck influence disentanglement in PitchFlower. Starting from the baseline with an RVQ module (8 levels, 512 codes each), we doubled the number of levels and compared against FSQ bottlenecks of similar capacity, using 9 quantization levels per dimension with 24, 46, and 90 dimensions. Figure~\ref{fig:rvq_fsq} shows a consistent gap between the two types: for a given capacity, RVQ yields better pitch controllability, suggesting stronger disentanglement. Increasing capacity in FSQ rapidly breaks controllability, while RVQ remains stable. As expected, smaller bottlenecks improve disentanglement for both.  

Finally, we tested a model without an autoencoder, where the flow directly reconstructs the original mel-spectrogram from its perturbed version. This system, lacking a bottleneck, failed to disentangle pitch, as the flow exploited residual information and WORLD artifacts to recover F0. These results confirm that perturbation and bottleneck play complementary roles: perturbation removes pitch cues explicitly, while the bottleneck prevents their recovery.

\subsection{Disentanglement vs Controllability}

We assess the question of whether models which exhibit higher degrees of controllability truly have better disentanglement. Our methodology is as follows: we extract the bottleneck features (RVQ or FSQ) from each trained model and attempt to predict the original pitch contour directly from these latent states. The prediction head has the same architecture as a decoder block, using a stride of 2 to match the temporal resolution of the F0 contour. It is tasked with simultaneously predicting two targets: a binary variable indicating whether a frame is voiced, and the corresponding log-F0 value for voiced frames. The head is optimized using binary cross-entropy and mean-squared-error losses, respectively.

\begin{table}[t]
    \centering
    \caption{Performance of pitch prediction head on top of each model's hidden states}
    \label{tab:pitch_probe}
    \begin{tabular}{lcc}
        \toprule
        \textbf{Model} & \textbf{Pitch Error (cents)} & \textbf{V/UV F1 Score} \\ 
        \midrule
        Bottleneck           & 232 & 0.96 \\
        \quad + HuBERT       & 197 & 0.96 \\
        \addlinespace
        Adversarial          & 348 & 0.94 \\
        \quad + HuBERT       & 230 & 0.96 \\
        \addlinespace
        PitchFlower          & 295 & 0.96 \\
        \quad + HuBERT       & 223 & 0.96 \\ 
        \midrule
        FSQ Model (90 dim)   & 98 & 0.96 \\ 
        \bottomrule
    \end{tabular}
\end{table}

As shown in Table~\ref{tab:pitch_probe}, even highly controllable models retain a non-trivial degree of pitch predictability within their latent codes. This residual predictability can be explained by the fact that these codes carry rich semantic and linguistic content, from which pitch can be implicitly inferred by the prediction head. Nonetheless, a clear pattern emerges: models that do not explicitly target pitch removal (either with adversarial approaches or perturbation) have a consistently smaller error, indicating higher retention of unwanted pitch information.

Interestingly, adding HuBERT distillation reduces pitch prediction error across all configurations. While a lower prediction error typically implies inferior disentanglement, here it does not compromise actual pitch controllability. In fact, as noted in Section~\ref{sub:disentanglement}, HuBERT generally improves F0 control reflected in lower F0RMSE curves. This indicates that the increased predictability is mediated by enforced semantic structures inside the codes, rather than leaked explicit F0 signals. Conversely, this stands in sharp contrast to the oversized FSQ baseline (90 dimensions), where the low prediction error reflects genuine acoustic leakage, causing its pitch controllability to rapidly break down as capacity increases.

\subsection{Flow inference parameters}
\begin{figure}[htb]
\begin{minipage}[b]{1.0\linewidth}
  \centering
  \includegraphics[width=\linewidth]{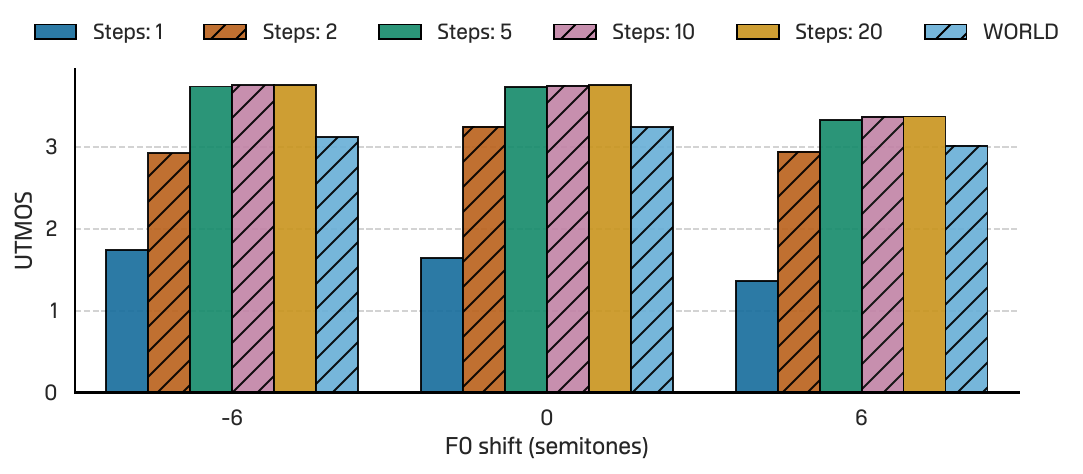}
  \centerline{(a)}
\end{minipage}
\begin{minipage}[b]{1.0\linewidth}
  \centering
  \centerline{\includegraphics[width=\linewidth]{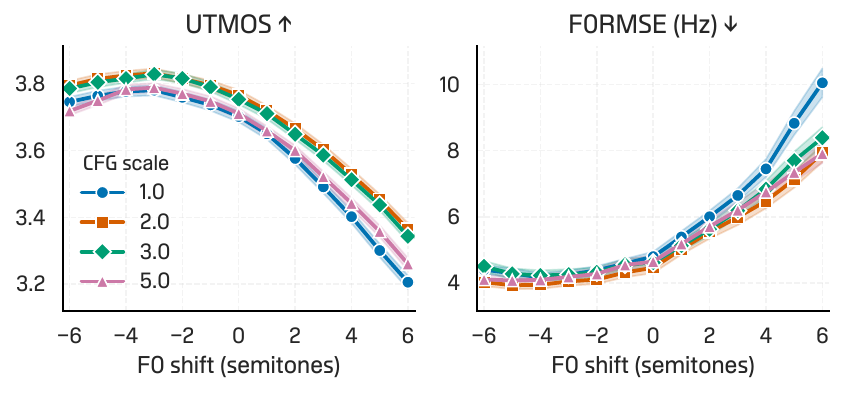}}
  \centerline{(b)}\medskip
\end{minipage}
\caption{(a) UTMOS score when changing the number of flow steps. (b) UTMOS and F0RMSE curves for different values of the CFG scale.}
\label{fig:flow_parameters}
\end{figure}

Since flow models require multiple evaluations to generate samples, we examine the effect of the number of steps. As shown in Figure~\ref{fig:flow_parameters}, audio quality (UTMOS) saturates at 5 steps, with no further improvements beyond this point. Notably, comparable quality to WORLD is already achieved with only 2 steps.  

We also study the classifier-free guidance scale. Figure~\ref{fig:flow_parameters} shows that increasing this parameter improves both audio quality and pitch controllability up to a limit. Values between 2.0 and 3.0 give the best performance, while 5.0 degrades audio quality with only marginal gains in controllability.  

\vspace{-0.3cm}

\vspace{0.5cm}

\section{Discussion}
\label{sec:discussion}

The foundational hypothesis of PitchFlower is that precise attribute control does not require complex, multi-stage optimization objectives; rather, it can be achieved through a simple perturbation framework. Fundamentally, attribute manipulation is a problem of representation disentanglement, where the goal is to isolate a target feature to allow for explicit conditioning. We argue that this isolation can be achieved by applying an input-level perturbation that specifically masks the target attribute. When such a perturbed signal is passed through an autoencoder tasked with reconstruction, and the model is simultaneously provided with the explicit, ground-truth attribute as a conditioning signal, disentanglement and control naturally emerge.

In the specific case of pitch, we employ a stochastic perturbation to deliberately break the correlation between the source audio and its original pitch trajectory. While we do not provide a formal mathematical proof, we hypothesize that this stochasticity is vital: a uniform, fixed pitch shift would present a predictable pattern that a powerful neural network could easily invert to recover the original F0. Furthermore, our empirical findings reveal that input perturbation alone is insufficient: an information bottleneck is required to prevent the model from exploiting residual pitch cues. Whether a bottleneck is required for other speech attributes will likely depend on a triad of factors: the inherent nature of the attribute, the strength of the perturbation, and model capacity. We hypothesize that highly entangled attributes, like emotion or timbre, may require even tighter, low capacity bottlenecks to enforce clean separation.

Another important finding of this work is the model's resilience to signal degradation during the perturbation phase. Although our input pitch modification was performed using WORLD, which inherently introduces vocoding artifacts and destroys fine acoustic details, PitchFlower successfully filtered out these imperfections during reconstruction. This demonstrates that a sufficiently capable generative decoder, such as our flow-matching module, can synthesize high-quality audio even when conditioned on a severely degraded input signal. This is a highly promising development for the generalizability of our framework: it implies that future extensions to more complex speech attributes can use aggressive, potentially destructive perturbations without sacrificing the final perceptual quality of the synthesized audio.

\subsection{Limitations}

The main limitation of our approach lies in the supported F0 range. Without inductive biases or explicit assumptions, the system can only generate values observed during training. Consequently, transposition quality depends on both the shift factor and the original F0. Shifts of up to $-1.5$ octaves are feasible when starting from high F0, but performance degrades at low frequencies, saturating near 60 Hz and failing to follow the contour. On LibriTTS, the effective range is roughly 60-700 Hz; extending beyond this requires training on data with a wider F0 distribution.  

Additionally, working with mel-spectrograms sets a clear upper limit to the quality of PitchFlower. Even with perfect reconstruction, the model is limited by the quality of the vocoder. Moreover, when pitch is modified, the resulting mel-spectrogram can have increased artifacts, which can be amplified by the vocoder and result in low quality audio. A clear direction to explore is making PitchFlower fully end-to-end. 

In a broader context, our framework relies on the assumption that a suitable perturbation targeting the desired attribute exists. While we successfully isolated pitch using the WORLD vocoder, defining an equivalent masking operation for other speech dimensions is not straightforward. For attributes like emotion or speaker identity, the underlying features are deeply entangled across time and frequency. Consequently, designing a perturbation that effectively masks the target attribute without entirely destroying the residual linguistic and acoustic content remains a significant challenge, potentially requiring substantial engineering.

\section{Conclusions}
\label{sec:conclusions}

We introduced PitchFlower, a neural audio codec that achieves pitch controllability through perturbation and information masking. Experiments show that this disentanglement strategy enables PitchFlower to surpass DSP-based baselines and perform on par with state-of-the-art neural approaches.

Our study of alternative disentanglement strategies highlights their different trade-offs. Bottleneck-only approaches achieve high fidelity but weak disentanglement. Adversarial methods and semantic distillation can further enhance disentanglement, yet at the cost of overall quality. Among these strategies, PitchFlower strikes the most favorable balance between pitch controllability and audio quality.  

Looking forward, the framework we propose is not limited to pitch. The same principles could be applied to other speech attributes such as emotion or timbre. We attribute PitchFlower’s effectiveness to three factors: (i) an effective perturbation that targets primarily the desired attribute, (ii) a bottleneck that prevents recovery and forces disentanglement, and (iii) a flow-based decoder capable of reconstructing realistic audio even from perturbed inputs. Together, these elements make PitchFlower a simple yet powerful step toward more controllable neural audio codecs.

\clearpage

\section*{Acknowledgments}
This work was partly performed using HPC resources from GENCI-IDRIS (Grant 2025-AD011016144), and funded by the ANR project EVA (ANR-23-CE23-0018).

\bibliographystyle{IEEEtran}
\bibliography{mybib}

\end{document}